\documentstyle[12pt]{article}
\begin{document}

\hfill DUKE-CGTP-99-03

\hfill hep-th/9902116

\vspace{1.5in}

\begin{center}

{\large\bf D-Branes, Derived Categories, }

{\large\bf and Grothendieck Groups }

\vspace{1in}

Eric Sharpe \\
Department of Physics \\
Box 90305 \\
Duke University \\
Durham, NC  27708 \\
{\tt ersharpe@cgtp.duke.edu} \\

 $\,$

\end{center}

In this paper we describe how Grothendieck groups of
coherent sheaves and locally free sheaves can be used
to describe type II D-branes, in the case that all
D-branes are wrapped on complex varieties and all
connections are holomorphic.  Our proposal is in the same
spirit as recent discussions of $K$-theory and D-branes;
within the restricted class mentioned, Grothendieck
groups encode 
a choice of connection on each D-brane worldvolume, in addition
to information about the $C^{\infty}$ bundles.
We also point out
that derived categories can also be used to give insight
into D-brane constructions, and analyze how a ${\bf Z}_2$
subset of the T-duality group acting on D-branes on tori
can be understood in terms
of a Fourier-Mukai transformation.

\begin{flushleft}
February 1999
\end{flushleft}

\newpage

\section{Introduction}

Recently it was noted that topological $K$-theory 
can be usefully employed to describe D-brane charges \cite{edktheory}.
In this paper we shall introduce new technical tools which
give a refinement of $K$-theory (more precisely, a holomorphic
version of $K$ theory), at the cost of less general
applicability.  As an example of their application, we will apply
these tools to a ${\bf Z}_2$ subgroup of T-duality (identified
with a Fourier-Mukai transform) and show how these tools can be used
to understand Fourier-Mukai transforms beyond the subclass of
sheaves usually considered in the physics literature.


It has been observed elsewhere (for example,
\cite{harveymoore,horioz}) that branes supported on
complex submanifolds of complex varieties are naturally described in terms
of coherent sheaves.  We shall describe
how Grothendieck groups of coherent sheaves, the holomorphic version
of $K$-theory referred to above, can be used to describe
D-branes, in the case that all D-branes are wrapped on complex
submanifolds.  Since we ultimately wish to study T-duality
realized as a Fourier-Mukai transform, and Fourier-Mukai transforms
are defined, in general, on derived categories, not individual
sheaves, we shall also discuss derived categories.
In particular, we shall point out a physical interpretation of
objects of a derived category, and give a physically-motivated map
from objects of a derived category to Grothendieck group elements.
We conclude with a discussion of 
T-duality symmetries in terms of Fourier-Mukai transforms.
In particular, we shall examine how Grothendieck groups can be
used to extend the action of Fourier-Mukai transforms beyond the
class of W.I.T. sheaves considered previously in the physics
literature.
For completeness, we have also included a short appendix on the 
basics of topological
$K$-theory.
We suspect the application of these technical tools
may have much broader applicability (to the study of Kontsevich's
mirror conjecture, for example), but unfortunately we shall
have little to say on such extensions.

We shall only consider D-branes in type II theories,
which are described by the $K$-theory of complex
vector bundles \cite{edktheory}.  We will not usually
work with space-filling D-branes, and so we shall not concern
ourselves with tadpole-cancellation issues.

In \cite{edktheory} it was noted that branes
can only consistently wrap a submanifold
when the normal bundle to the submanifold
admits a $\mbox{Spin}^c$ structure.
In this paper we will only work in complex
geometry, and as all $U(N)$ bundles
admit a canonical $\mbox{Spin}^c$ structure
\cite[appendix D]{lm}, we shall never have
to consider this subtlety in this paper.
(For a more thorough discussion of the $\mbox{Spin}^c$
constraint in the context of type II compactifications
with vanishing cosmological constant, see \cite{robme}.)

We shall assume throughout this paper that all
varieties are smooth and projective.
(For example, all complex tori appearing will
implicitly be assumed to be abelian varieties.)

Since the publication of \cite{edktheory},
several other papers have appeared on topological
$K$ theory and D-branes \cite{petr,hugo,sergei,pyi,berggaber}.
We should also mention that the work \cite{edktheory}
built upon the earlier works \cite{sen1,sen2,sen3,sen4,gregminas}.
We have also been informed that another discussion of T-duality
in the context of $K$-theory will appear in \cite{petroreneric}.
Also, as this paper was being finalized, another paper on
T-duality and $K$-theory appeared \cite{horinew}.

\section{Grothendieck groups}

It has recently been argued by E. Witten that D-brane charges
should be understood in terms of topological $K$-theory
\cite{edktheory}.  In this paper, we shall argue that
in certain cases it is more useful to work with Grothendieck
groups of coherent sheaves.  In this particular section we shall
define Grothendieck groups, then in later sections we shall show
their relation to derived categories and describe how they can
give insight into a ${\bf Z}_2$ subgroup of T-duality realized
as a Fourier-Mukai transformation.

In this paper we shall only work on
complex varieties, and will only wrap branes on (complex)
subvarieties.  This constraint reduces us to a proper subset of all 
possible D-brane configurations, but by making this restriction we 
will be able to use more powerful tools.  For example, in these circumstances 
we can make some
strong statements concerning supersymmetric vacuum configurations
of a D-brane \cite[section 4.2]{harveymoore}.  Consider
a set of $N$ branes on some K\"ahler variety of dimension $n$. 
If $F$ is the curvature of the connection on the $U(N)$ bundle,
and $J$ the K\"ahler form,
then in order to get a supersymmetric vacuum some necessary 
conditions\footnote{The attentive reader will note that these are 
almost, but not quite,
two necessary conditions for supersymmetric heterotic vacua.}
on $F$ are \cite[section 4.2]{harveymoore}
\begin{eqnarray}
F & \in & \Omega^{1,1} \label{holcond} \\
F \wedge J^{n-1} & = & \lambda J^{n}  \label{duyanalogue}
\end{eqnarray}
for some constant $\lambda$.

Given a $C^{\infty}$ bundle ${\cal E}$ (with a fixed Hermitian structure)
on a complex manifold,
there is a one-to-one correspondence between connections $D_A$ on ${\cal E}$
that satisfy equation~(\ref{holcond})
(in other words, holomorphic connections) and holomorphic structures
on ${\cal E}$ \cite[section VII.1]{kobayashi}.  Thus, specifying a bundle
with a fixed holomorphic structure is equivalent to specifying a 
$C^{\infty}$ bundle with a choice of holomorphic connection.
(If in addition the holomorphic connection satisfies  
equation~(\ref{duyanalogue}), then the corresponding holomorphic
bundle will be Mumford-Takemoto semistable.)

Thus, within the context of the restriction to complex subvarieties
and holomorphic bundles, the specification of a holomorphic bundle
on some subvariety is equivalent to specifying a complex $C^{\infty}$ bundle
together with a choice of holomorphic connection on the bundle -- data
associated
with a D-brane.  

Instead of working with topological $K$ theory, 
which only encodes $C^{\infty}$ bundles,
it can be advantageous to work with a ``holomorphic'' version of $K$-theory,
which implicitly encodes not only choices of $C^{\infty}$ bundles,
but also specific choices of (holomorphic) connections on the bundles.
Such a holomorphic version of topological $K$-theory exists,
and is known as a Grothendieck group (of locally free sheaves).

Before we actually define Grothendieck groups, we need to make
some general observations.
The motivation given above
for working with Grothendieck groups is clearly rather weak,
but in later sections we shall give stronger arguments.  
We pointed out that the conditions for a supersymmetric
D-brane vacuum on a complex K\"ahler manifold imply that the connection
on the $C^{\infty}$ bundle is holomorphic, and so the combined $C^{\infty}$
bundle plus connection can be described equivalently in terms of a 
holomorphic bundle.
However, when we start working with configurations of 
both branes and antibranes, we should not expect conditions for
a supersymmetric vacuum to be of great relevance, and so it is not
completely clear from this description that Grothendieck groups are
necessarily useful objects.
We shall see later that working with Grothendieck groups give us
a natural arena in which to examine T-duality, for example, so by
working with Grothendieck groups we do get some useful insights.

Before defining Grothendieck groups, another technical observation should
be made.  In order to specify a
supersymmetric vacuum for a D-brane, we must specify not just any holomorphic
connection, but one which is Hermitian-Einstein (equation~(\ref{duyanalogue})). 
Thus, to specify a supersymmetric vacuum, not any holomorphic bundle
will do, but only those which are Mumford-Takemoto
semistable.
Note this means that given a general element of the 
Grothendieck group, there is not one but two reasons why it
will not describe a supersymmetric vacuum -- not only because
of the simultaneous presence of branes and antibranes, but also because
the (holomorphic) bundles are not necessarily Mumford-Takemoto semistable.

Strictly speaking there are two distinct Grothendieck
groups relevant here, which we shall denote $K'^0(X)$ and $K'_0(X)$
\cite{manin,sga6}.  We shall first define both, then point out
that in reasonably nice circumstances they are isomorphic.
To distinguish Grothendieck groups from topological
$K$-theory, we shall use $K'(X)$ to denote
Grothendieck groups and $K(X)$ to denote topological $K$-theory.

The Grothendieck group $K'_0(X)$ of coherent sheaves is
defined to be the free abelian group on
coherent sheaves on $X$, modulo elements
${\cal E} - {\cal E}' - {\cal E}''$,
where ${\cal E}$, ${\cal E}'$, and ${\cal E}''$
are coherent sheaves related by short exact sequences of the form
\begin{displaymath}
0 \rightarrow {\cal E}'' \rightarrow {\cal E}
\rightarrow {\cal E}' \rightarrow 0
\end{displaymath}

The Grothendieck group $K'^0(X)$ of locally free sheaves
is defined to be the free abelian group on locally free sheaves
on $X$, modulo elements ${\cal E} - {\cal E}' - {\cal E}''$,
where ${\cal E}$, ${\cal E}'$, and ${\cal E}''$ are locally
free sheaves related by short exact sequences of the form
\begin{displaymath}
0 \rightarrow {\cal E}'' \rightarrow {\cal E}
\rightarrow {\cal E}' \rightarrow 0
\end{displaymath}
More formally \cite[prop. 4.4]{manin}, $K'^0$ is a contravariant
functor from the category of noetherian schemes to the category of rings.
Note in passing that these definitions of $K'_0$ and $K'^0$ are 
closely analogous to the definition of topological $K^0$.

For further information on Grothendieck groups (and their relation
to derived categories, which shall appear shortly), 
see for example \cite{manin,sga6}.

It can be shown (\cite{manin},\cite[exercise III.6.9]{hartshorne})
that on a smooth projective
variety $X$, the natural map $K'^0(X) \rightarrow K'_0(X)$
is an isomorphism.  In the rest of this paper we shall assume
that we are always working on a smooth projective variety,
and so we shall use $K'^0$ and $K'_0$ more or less
interchangeably.  We shall also often refer to ``the'' 
Grothendieck group.

The reader may wonder how precisely Grothendieck groups are related
to topological $K$-theory.
In order to get some insight into the relation between these objects,
let us consider an example.
Suppose $X$ is a smooth compact Riemann surface.
It is straightforward to compute\footnote{Using the Atiyah-Hirzebruch spectral
sequence.  See \cite[section 2]{ah}.} that topological $K^0(X) = {\bf Z}^{
\oplus 2}$.  By contrast \cite[exercise II.6.11]{hartshorne}, the 
Grothendieck group $K'_0(X) = 
\mbox{Pic } X \oplus {\bf Z}$.
Although the topological $K$-theory groups and Grothendieck groups
are not identical, they are still closely related.
Note for example that for $X$ a smooth Riemann surface,
$\mbox{Pic }X$ is an extension of ${\bf Z}$ by $\mbox{Jac }X$,
so the Grothendieck group $K'_0(X) = \mbox{Pic }X \oplus {\bf Z}$
includes the topological $K$-theory group $K^0(X) = {\bf Z} \oplus {\bf Z}$
as a subset.  In other words, in this example the Grothendieck
group contains more information than topological $K^0$.
This certainly agrees with the intuition we laid earlier -- the
Grothendieck group should contain information not only about
the choice of $C^{\infty}$ bundle, but also about the precise choice
of connection on that bundle.

In general it is easy to see that the Grothendieck group $K'^0$
maps into topological $K^0$.  Unfortunately in general this
map will not be surjective.
One can certainly map a locally free sheaf to a smooth bundle,
essentially just by forgetting the holomorphic structure.
The attentive reader might be concerned that this map
is not well-defined -- in the
definition of $K'^0$,
${\cal E}$ is identified with ${\cal E}' \oplus {\cal E}''$
if ${\cal E}$ is an extension of either ${\cal E}'$ or ${\cal E}''$
by the other, whereas in topological $K^0$ we only identify split
extensions.  However, it is a standard fact that any 
extension of continuous vector bundles splits
\cite[section 3.9]{husemoller} (whereas not every
extension of holomorphic bundles splits holomorphically),
so in fact the obvious map $K'^0 \rightarrow K^0$
is well-defined.  Unfortunately in general this map will not
be surjective.  One way to see this is to note that Chern classes
of a holomorphic bundle on a projective variety $X$ live only in
a subset of $H^*(X,{\bf Z})$ -- in particular,
$c_i \in H^{(i,i)}(X) \cap H^{2i}(X,{\bf Z})$ -- whereas
Chern classes of an arbitrary $C^{\infty}$ complex bundle are not so restricted.

In topological $K$-theory, one can define $K^1$ in addition to $K^0$.
There are also holomorphic versions of $K^1$, though they are
rather more obscure \cite[chapter 13]{swanbook}.  We shall not use
these holomorphic versions of $K^1$, though for completeness we
list them here.

Define \cite[chapter 13]{swanbook}
$K'_1(X)$ to be the free abelian group on pairs $({\cal E}, \rho)$
where ${\cal E}$ is a coherent sheaf on $X$ and $\rho: {\cal E}
\rightarrow {\cal E}$ is an isomorphism,
modulo elements $({\cal E}, \rho) - ({\cal E}', \rho') - ({\cal E}'',
\rho'')$ where ${\cal E}$, ${\cal E}'$, and ${\cal E}''$ are
coherent sheaves related by short exact sequences of the form
\begin{displaymath}
0 \rightarrow {\cal E}'' \rightarrow {\cal E} \rightarrow {\cal E}' 
\rightarrow 0
\end{displaymath}
and also modulo $({\cal E}, \rho \circ \psi) - ({\cal E}, \rho)
- ({\cal E}, \psi)$.

Define \cite[chapter 13]{swanbook}
$K'^1(X)$ analogously to $K'_1(X)$, that is,
to be the free abelian group on pairs $({\cal E}, \rho)$
where ${\cal E}$ is a locally free sheaf on $X$ and $\rho: {\cal E}
\rightarrow {\cal E}$ is an isomorphism,
modulo elements $({\cal E}, \rho) - ({\cal E}', \rho') - ({\cal E}'',
\rho'')$ where ${\cal E}$, ${\cal E}'$, and ${\cal E}''$ are
locally free sheaves related by short exact sequences of the form
\begin{displaymath}
0 \rightarrow {\cal E}'' \rightarrow {\cal E} \rightarrow {\cal E}' 
\rightarrow 0
\end{displaymath}
and also modulo $({\cal E}, \rho \circ \psi) - ({\cal E}, \rho)
- ({\cal E}, \psi)$.

In passing, note that these definitions are closely analogous
to a definition of topological $K^1$ used recently in,
for example, \cite{petr}.

\section{Derived categories}      \label{drcatintro}

Ultimately in this paper we would like to study the action of
T-duality (realized as a Fourier-Mukai transform) on
brane/antibrane configurations.  However, Fourier-Mukai transforms
are defined on derived categories of coherent sheaves, not
individual sheaves, in general.  In special cases\footnote{Indeed,
previously in the physics literature authors have only considered  
these special cases when discussing Fourier-Mukai transforms.} one can 
make sense out of the action of a Fourier-Mukai transform
on an individual sheaf, however to discuss Fourier-Mukai transforms
in generality, one must turn to derived categories.

Because of our interest in T-duality, 
we shall now discuss derived categories
and their physical relevance.  In particular, we shall show how an element of a
Grothendieck group can be obtained from an object in a derived
category (in a physically meaningful manner).  In the next section,
we shall put this map to use in studying T-duality in terms of
Fourier-Mukai transformations.
As usual, we shall be implicitly
working over complex varieties and with holomorphic bundles,
and so we shall also assume that all tachyons, viewed as bundle maps,
are also complex and holomorphic.

Recall from \cite{edktheory} that given a
coincident brane, anti-brane pair, with bundles
${\cal E}$ and ${\cal F}$ respectively,
and a tachyon field $T: {\cal E} \rightarrow {\cal F}$,
then the resulting brane charge one would actually
be left with in vacuum is (at least morally) the Grothendieck group element
given by $\ker T \ominus \mbox{coker } T$,
or equivalently $H^0 \ominus H^1$,
where the $H^i$ are the cohomology of the complex
\begin{displaymath}
0 \longrightarrow {\cal E} \stackrel{T}{\longrightarrow}
{\cal F} \longrightarrow 0
\end{displaymath}
(Note that since we are working in complex geometry, $\ker T$ and 
$\mbox{coker }T$
make sense as sheaves\footnote{A technical note:  we shall 
implicitly restrict to complexes whose cohomology sheaves
are coherent.}.)  

One can also imagine working with more general
complexes of bundles.  These would be described
as a sandwich of alternating branes and anti-branes.
For example, let ${\cal F}^{\bullet}$ denote a
complex of bundles 
\begin{displaymath}
\cdots \stackrel{T_{i-1}}{\longrightarrow} 
{\cal F}^{i} \stackrel{T_{i}}{\longrightarrow}
{\cal F}^{i+1} \stackrel{T_{i+1}}{\longrightarrow}
{\cal F}^{i+2} \stackrel{T_{i+2}}{\longrightarrow} \cdots
\end{displaymath}
(where, by definition of complex\footnote{
Note that, for example, $T_{2j+1} \circ T_{2j}$ is a map from the total
sheaf on the brane (equation~(\ref{totbrane})) back into itself, whereas
tachyons should only map branes to antibranes and vice-versa.  Thus,
in order to consistently break up the tachyon between the total
brane (\ref{totbrane}) and antibrane (\ref{totantibrane}) into an 
interweaving series of maps, as will be mentioned shortly, we must demand
that $T_{j+1} \circ T_j = 0$, i.e., that the maps define a complex.}, 
$T_{j+1} \circ T_j = 0$)
such that the ${\cal F}^{2i}$
all live on (coincident) branes and the
${\cal F}^{2i+1}$ all live on (coincident) anti-branes.
Put another way, the total sheaf on the brane is
\begin{equation}  \label{totbrane}
\bigoplus_n {\cal F}^{2n}
\end{equation}
and the total sheaf on the antibrane is
\begin{equation}  \label{totantibrane}
\bigoplus_n {\cal F}^{2n+1}
\end{equation}
with the tachyon potential broken up into an interweaving
series of maps between the brane and antibrane.
After cancelling as much as possible,
one is left with an element of $K'_0(X)$ given by
\begin{displaymath}
\left[ \bigoplus_{n } H^{2n} \right]
\ominus \left[ \bigoplus_{n } H^{2n+1} \right]
\end{displaymath}

Clearly such a complex encodes a lot of physically-irrelevant
information.  Indeed, the complex ${\cal F}^{\bullet}$ described
above is physically identical to the complex
\begin{displaymath}
\cdots \stackrel{0}{\longrightarrow} H^{j-1} \stackrel{0}{\longrightarrow}
H^j \stackrel{0}{\longrightarrow} H^{j+1} \stackrel{0}{\longrightarrow}
\cdots
\end{displaymath}
and even the complex
\begin{displaymath}
0 \longrightarrow \bigoplus_{n} H^{2n} \stackrel{0}{\longrightarrow}
\bigoplus_{n} H^{2n+1} \longrightarrow 0
\end{displaymath}
The only physically relevant aspect of the complex is its
image in the Grothendieck group.

Although the only physically relevant aspect of a complex of
branes and anti-branes is its Grothendieck-group image, we shall
nevertheless find it useful to work in terms of complexes
in the next section.

Since we are working in algebraic
geometry, the attentive reader may wonder why we are restricting to locally
free sheaves, rather than considering general coherent sheaves.
For example, we could identify a torsion sheaf with a lower-dimensional
D-brane.  The difficulty is that we wish to speak of maps between
the worldvolumes described by tachyons, and although open strings
connecting branes and antibranes of the same dimension certainly
contain tachyon modes, open strings connecting branes and antibranes
of distinct dimension need not contain tachyon modes -- whether
a tachyon is actually present varies from case to case.  Thus, 
we are restricting to locally free sheaves on worldvolumes
all of the same dimension.

There exists a useful mechanism for working
with complexes of holomorphic bundles, and more generally,
holomorphic sheaves.  This tool is known as a derived category.

A derived category of coherent sheaves on some
variety $X$ is a category whose objects
are complexes of sheaves on $X$, such that the
cohomology sheaves of the complexes are coherent.
A derived category of (bounded complexes of) 
sheaves on $X$ is denoted $D^b(X)$.
The subcategory defined by complexes of sheaves with coherent
cohomology is denoted $D^b_c(X)$.
In general, not all derived categories are derived categories
of sheaves; however, all the derived categories we shall
describe in this paper are derived categories of sheaves.

A proper explanation of derived categories is well beyond
the scope of this paper -- for more information, see
for example \cite{weibel,hartRD}.  However we shall mention
one useful fact in passing.  Morphisms of chain complexes that
preserve cohomology (so-called quasi-isomorphisms) descend to
isomorphisms in the derived category, so intuitively the reader might,
very loosely\footnote{Technically this is incorrect;
however, for those readers unwilling to delve into technicalities,
this description does give some handle on matters.}, 
imagine that any two complexes in the derived category
with isomorphic cohomology groups are themselves considered isomorphic.

It has been speculated previously in the physics
literature that derived categories were relevant
for physics \cite{paulron,zaslow,kont}.
In the context of holomorphic bundles, we now have
an explicit correspondence.

Note that derived categories, just like complexes, 
contain a great deal of physically irrelevant information.
We do not need to know the full cohomology of a complex
of sheaves, but only the formal difference
\begin{displaymath}
\left( \bigoplus_n H^{2n} \right) \ominus
\left( \bigoplus_n H^{2n+1} \right)
\end{displaymath}
In other words, the only physically relevant part of an
object in a derived category is its image in the Grothendieck
group of coherent sheaves.

The attentive reader should be slightly bothered by our
use of derived categories to describe sandwiches of branes and
antibranes.  In these brane/antibrane sandwiches, we implicitly
assumed that all the branes and antibranes were of the same dimension
(equivalently, that we had a locally free sheaf on the worldvolume
of each).  By contrast, the objects of a derived category 
are complexes of more or less arbitrary sheaves, whose
cohomology groups are coherent sheaves.  Naively, it would
seem that our brane/antibrane sandwich construction can only
sense a small portion of the possible objects of a derived category.

However, this is not the case.  Any bounded complex of coherent
sheaves on a smooth variety is quasi-isomorphic to a bounded
complex of locally free sheaves, that is, admits a chain map
to a complex of locally free sheaves such that the chain map
preserves the cohomology of the complex.  (This is known formally
as a Cartan-Eilenberg resolution of the complex \cite[section 5.7]{weibel}.)
Since quasi-isomorphisms descend to isomorphisms in the derived category,
we see that any complex of coherent sheaves is isomorphic
(within the derived category) to a complex of locally free sheaves.

There is one further technical problem that might bother
the attentive reader.  We have just argued that any complex
of coherent sheaves can be equivalently described by a complex
of locally free sheaves, and so in terms of a brane/antibrane
sandwich.  However, the objects of a derived category
are not precisely complexes of coherent sheaves, but rather
complexes of sheaves whose cohomology sheaves are coherent.
In the special case of sheaves on smooth projective varieties,
we strongly suspect that the two categories are equivalent,
but we do not have a rigorous argument to support that claim.

\section{T-duality}

Now that we have introduced relevant technical machinery,
we shall discuss T-duality.  It is often said that a ${\bf Z}_2$
subgroup of T-duality
is realized via Fourier-Mukai transforms \cite{horioz}, 
and in the present context
we shall find a natural setting for this ansatz.
We shall begin by giving a physical motivation for the identification
of a ${\bf Z}_2$ subgroup of T-duality with a Fourier-Mukai
transform, then go through a number of technical results
on Fourier-Mukai transforms, and finally conclude with a discussion
of why precisely one needs Grothendieck groups and derived categories to
discuss Fourier-Mukai transforms on general D-brane configurations.

\subsection{Physical motivation}

Before we begin discussing Fourier-Mukai transforms in technical
detail, we shall discuss in a pair of examples why precisely
it is sometimes claimed \cite{horioz} that a ${\bf Z}_2$ subgroup
of the T-duality group acting on branes wrapped on complex algebraic
tori is realized as a Fourier-Mukai transform.
Note that
for D-branes wrapped on $T^{2g}$, when we speak of T-duality
we mean, T-duality along each of $2g$ $S^1$'s in $T^{2g}$.  Since
we have T-dualized an even number of times, we will always take
type IIA back to type IIA, and type IIB back to type IIB. 

1)  Consider a rank $N$ bundle on $T^2$, with $c_1 = 0$ -- in other words,
an $SU(N)$ bundle on $T^2$.  This precisely corresponds
to $N$ $Dp$-branes wrapped on $T^2$, with no immersed $D(p-2)$-brane
charge.  One expects that T-duality should map this to a configuration
of $N$ $D(p-2)$-branes, with support only at points on $\hat{T}^2$.

Indeed, this is precisely what we find.  For reasonably nice\footnote{In
notation to be defined shortly, W.I.T.$_1$.} $SU(N)$ bundles ${\cal E}$
on $T^2$, the Fourier-Mukai transform is a skyscraper sheaf on $\hat{T}^2$,
supported at points.

2)  Consider a rank $N$ bundle ${\cal E}$ on $T^4$ -- in
other words, a $U(N)$ bundle on $T^4$.  This precisely corresponds
to $N$ $Dp$-branes wrapped on $T^4$, with immersed $D(p-2)$-brane
charge $c_1({\cal E})$, 
and with $D(p-4)$-brane charge given by $ch_2({\cal E}) = 
c_2({\cal E}) - (1/2) c_1({\cal E})^2$.  Under T-duality
we expect $D(p-4)$-brane charge\footnote{A small clarification
is in order.  Given some $Dp$-brane, there are two ways to get,
say, $D(p-4)$-brane charge:  (i) add a $D(p-4)$-brane (add a torsion
sheaf, in more algebraic language), and (ii) modify $ch_2$ of the bundle
on the $Dp$-brane worldvolume.  More globally one expects the moduli
space to be more or less reducible, with these options corresponding
to distinct components.  For simplicity we only discuss option (ii)
in the example above.} on $T^4$ to become
$Dp$-branes wrapped $\hat{T}^4$, and $Dp$-branes wrapping $T^4$ to become
$D(p-4)$-brane charge on $\hat{T}^4$.  Thus, we expect
the T-dual to this configuration to be another bundle $\hat{{\cal E}}$
on the dual $T^4$,
of rank $ch_2({\cal E})$, and $ch_2(\hat{{\cal E}}) = \mbox{rank }{\cal E}$.

Indeed, this is precisely what we find.  For reasonably nice\footnote{In
notation to be defined shortly, W.I.T.$_1$.} bundles ${\cal E}$ on $T^4$,
the dual is a bundle $\hat{{\cal E}}$ of\footnote{The equations shown
correct typographical errors in equation~(3.2.16) of \cite{donkron}.
We would like to thank Kentaro Hori for pointing out these errors
to us.}
\begin{eqnarray*}
\mbox{rank }\hat{{\cal E}} & = & ch_2({\cal E}) \\
  & = & c_2({\cal E}) - (1/2) c_1({\cal E})^2 \\
c_1(\hat{{\cal E}}) & = & \sigma( \, c_1({\cal E}) \, ) \\
ch_2(\hat{{\cal E}}) & = & \mbox{rank }{\cal E}
\end{eqnarray*}
where $\sigma: H^2(T^4, {\bf Z}) \stackrel{\cong}{\longrightarrow}
H^2(\hat{T}^4, {\bf Z})$ is an isomorphism.

Thus, at least in these two examples, the usual claim \cite{horioz} that
T-duality of branes is realized by Fourier-Mukai transform seems
to check out.

In discussions of Fourier-Mukai transforms in the physics literature,
a single sheaf is mapped to a single sheaf.  This is not 
the most general way that Fourier-Mukai transforms act;
it is also not the most natural.  In general, Fourier-Mukai transforms
act on derived categories of coherent sheaves, that is, they
act on complexes of sheaves.  One can act on a single sheaf ${\cal E}$ by
using the trivial complex
\begin{displaymath}
0 \rightarrow {\cal E} \rightarrow 0
\end{displaymath}
but in general the Fourier-Mukai transform will not be another trivial
complex, but a much more complicated complex.  In the next section
we shall give the general technical definition of a Fourier-Mukai transform,
then describe the special cases in which it has a well-defined action
on individual coherent sheaves.

\subsection{Technical definitions}

Let $X$ and $\hat{X}$ be projective varieties (not necessarily
tori, for the moment).
A Fourier-Mukai transform is a functor ${\cal T}$
between (in fact, an equivalence of) the derived categories
$D^b(X)$ and $D^b(\hat{X})$.  More precisely, if $\pi_1: X \times \hat{X} 
\rightarrow X$
and $\pi_2: X \times \hat{X} \rightarrow \hat{X}$ are the obvious projections,
then for any ${\cal P} \in \mbox{Ob }D^b(X \times \hat{X})$,
we can define a Fourier-Mukai functor\footnote{In general,
any equivalence of derived categories $D^b(X)$ and
$D^b(\hat{X})$ for any smooth projective varieties $X$ and $\hat{X}$ can 
be written in the
form of equation~(\ref{genfmdef})
for some ${\cal P} \in \mbox{Ob } D^b(X \times \hat{X})$ \cite{orlov}.}
\cite{mukai1}
\begin{equation}   \label{genfmdef}
\underline{R} \pi_{2 *}\left( {\cal P} \stackrel{{\bf L}}{\otimes} \pi_1^* - 
\right):
D^b(X) \rightarrow D^b(\hat{X})
\end{equation}
In the special case that ${\cal P}$ is a locally free sheaf on
$X \times X'$ (the only case we shall consider), the Fourier-Mukai
functor simplifies to become the right-derived functor\footnote{
As an aside, it is perhaps worth mentioning that conditions for
a locally free sheaf ${\cal P}$ to define an equivalence of
categories via equation~(\ref{fmdef}) are known \cite{bridge}.
The locally free sheaf ${\cal P}$ defines an equivalence
of categories via equation~(\ref{fmdef}) precisely when
for all points $x \in X$, ${\cal P}_x$ is simple, ${\cal P}_x
= {\cal P}_x \otimes \omega_{\hat{X}}$ (where $\omega_{\hat{X}}$ is 
the dualizing
sheaf on $\hat{X}$), and for any two distinct points $x_1$, $x_2$ of $X$
and any integer $i$, one has  $\mbox{Ext}^i_{\hat{X}}({\cal P}_{x_1},
{\cal P}_{x_2}) = 0$ \cite{bridge}.}
\begin{equation}   \label{fmdef}
\underline{R} \pi_{2 *}( {\cal P} \otimes \pi_1^* - ):
D^b(X) \rightarrow D^b(\hat{X})
\end{equation}
We shall denote this functor by 
${\cal T}: D^b(X) \rightarrow D^b(\hat{X})$, and we shall usually
restrict to $D^b_c(X)$ and $D^b_c(\hat{X})$.

In the remainder of this section, we shall specialize to the case that
$X$ and $\hat{X}$ are dual projective complex tori, and that
${\cal P}$ is the Poincare bundle on $X \times \hat{X}$.

Although Fourier-Mukai transforms are defined on derived categories,
that is, on complexes of sheaves, there is a way to make sense
out of their action on individual sheaves in special cases,
and this is the specialization usually invoked in the physics
literature.  First, note that given any coherent sheaf ${\cal E}$,
we can define the trivial complex 
\begin{equation}  \label{trivcpx}
0 \rightarrow {\cal E} \rightarrow 0
\end{equation}
thus we can map individual coherent
sheaves into the class of objects of a derived category.
We say a coherent sheaf ${\cal E}$ is W.I.T.$_n$ if \cite{mukai1} 
\begin{displaymath}
R^i \pi_{2 *}\left( {\cal P} \otimes \pi_1^* {\cal E}\right) \: = \: 0
\end{displaymath}
for all $i$ except $i = n$.  Then, the Fourier-Mukai transform of
a sheaf ${\cal E}$, identified with an object of the derived category
via the trivial complex~(\ref{trivcpx}), is another sheaf
(also defined via~(\ref{trivcpx})), given by
\begin{displaymath}
\hat{{\cal E}} \: = \: R^n \pi_{2 *} \left( {\cal P} \otimes \pi_1^* 
{\cal E} \right)
\end{displaymath}
Moreover, it can be shown that if ${\cal E}$ is W.I.T.$_n$ for some $n$,
then $\hat{ {\cal E} }$ is also W.I.T.$_{n'}$ for some $n'$,
and moreover $\hat{ \hat{ {\cal E} } } = (-1)^*{\cal E}$,
where $(-1)$ multiplies all coordinates on the torus by $-1$ \cite{mukai1}.
Clearly, those coherent sheaves that are W.I.T.$_n$ for some $n$
have well-behaved dualization properties, and so physicists
speaking of Fourier-Mukai transformations usually assume the
sheaves in question are all W.I.T.
For example, in the examples at the beginning of this section,
it was assumed that the coherent sheaves given were W.I.T.$_1$.
However, not all coherent sheaves of interest are W.I.T.,
and for the more general case one needs the more general
methods outlined in this paper.  We shall speak to the more
general case, and the precise relevance of the W.I.T. condition,
in a later section.

\subsection{Action on Grothendieck groups}

Although Fourier-Mukai transforms are defined on derived categories,
they factor into an action on Grothendieck groups of coherent
sheaves, in a manner
that should be suggested by the physical setup of section~\ref{drcatintro}.
Let $\alpha_X: D^b_c(X) \rightarrow K'_0$ be defined
as the map that takes a complex of sheaves into the alternating
sum of the cohomologies of the complex, i.e.,
\begin{displaymath}
\alpha_X: {\cal F}^{\bullet} \rightarrow \left[
\oplus_n H^{2n}({\cal F}^{\bullet}) \right] \ominus
\left[ \oplus_n H^{2n+1}({\cal F}^{\bullet}) \right]
\end{displaymath}
(the same map we introduced in more physical terms in section~\ref{drcatintro})
and let ${\cal T}: D^b_c(X) \rightarrow D^b_c(\hat{X})$ denote
Fourier-Mukai transform, then we have a commutative diagram
\cite{beauville1}
\begin{displaymath}
\begin{array}{ccc}
\mbox{Ob }D^b_c(X) & \stackrel{\alpha_X}{\longrightarrow} & K'_0(X) \\
{\cal T} \downarrow & & \downarrow {\cal T}_K \\
\mbox{Ob }D^b_c(\hat{X}) & \stackrel{\alpha_{\hat{X}}}{\longrightarrow} & 
K'_0(\hat{X})
\end{array}
\end{displaymath}
where ${\cal T}_K: K'_0(X) \rightarrow K'_0(\hat{X})$ is 
defined by
\begin{displaymath}
{\cal T}_K(-) \: = \: \sum_i (-)^i R^i\pi_{2 *}( {\cal P} \otimes \pi_1^* -)
\end{displaymath}
In other words, the action of Fourier-Mukai transforms on
derived categories factors into an action on Grothendieck groups
of coherent sheaves.

\subsection{A sign ambiguity}

The attentive reader will notice there is a minor sign
ambiguity in our presentation of Fourier-Mukai transformations.
One typically defines the inverse of a Fourier-Mukai transformation
with minor sign asymmetries relative to the original transformation
\cite[section 3.2]{donkron}, just as inverses of Fourier transformations
are often defined with relative signs.  By contrast, we have presented
Fourier-Mukai transformations in an implicitly symmetric fashion,
which means our results can only be interpreted physically
up to a ${\bf Z}_2$ ambiguity.

In order to describe this sign problem more precisely, let us
reconsider the two examples given at the beginning of the section,
being somewhat more careful about signs.

1)  Consider a holomorphic rank $N$ bundle ${\cal E}$ on $T^2$, 
with $c_1 = 0$ -- in other
words, an $SU(N)$ bundle on $T^2$.  As mentioned earlier,
we assume ${\cal E}$ is W.I.T.$_1$, so
\begin{displaymath}
R^0 \pi_{2 *}( {\cal P} \otimes \pi_1^* {\cal E}) \: = \: 0
\end{displaymath}
A close examination of our definition of Fourier-Mukai transform reveals
that, as an element of $K'_0(\hat{T}^2)$, the Fourier-Mukai transform
of ${\cal E}$ is not precisely the torsion sheaf
$\hat{{\cal E}} = R^1 \pi_{2 *} \left( {\cal P} \otimes \pi_1^* {\cal E} \right)$
but rather the virtual torsion sheaf $\ominus \hat{ {\cal E} } \in
K'_0( \hat{T}^2 )$.

2)  Consider a holomorphic rank $N$ bundle ${\cal E}$ on 
$T^4$ -- in other words,
a $U(N)$ bundle on $T^4$.  Assume the complex structure on $T^4$ is
such that the $T^4$ is projective.
Earlier we mentioned that the Fourier-Mukai transform of ${\cal E}$
is a bundle $\hat{{\cal E}}$ on $\hat{T}^4$, in the case that
${\cal E}$ is W.I.T.$_1$, namely
\begin{eqnarray*}
R^0 \pi_{2 *}( {\cal P} \otimes \pi_1^* {\cal E}) & = & 0 \\
R^2 \pi_{2 *}( {\cal P} \otimes \pi_1^* {\cal E}) & = & 0
\end{eqnarray*}
A close examination of our definition of the Fourier-Mukai transform
reveals that, as an element of $K'_0( \hat{T}^4 )$, the Fourier-Mukai
transform of ${\cal E}$ is not precisely the bundle $\hat{{\cal E}} =
R^1 \pi_{2 *}( {\cal P} \otimes \pi_1^* {\cal E})$
but rather the virtual bundle $\ominus \hat{{\cal E}} \in K'_0( \hat{T}^4)$.

As it has been presented so far, this sign problem could naively be cured by
redefining the Fourier-Mukai transform.  Unfortunately, the difficulty
is much deeper.
Consider applying a Fourier-Mukai transform twice.
If we are studying branes wrapped on $T^{2g}$, then this means
T-dualizing along each of the $2g$ $S^1$'s in $T^{2g}$ twice,
and so intuitively we should return to where we started.
According to \cite{mukai1}, ${\cal T}^2 = (-1)^* [-g]$
as an action on $D^b(T^{2g})$.  The $[-g]$ formally shifts
all complexes g places to the right, and the $(-1)$ multiplies
all complex coordinates on the torus by $-1$.  This descends
to an action on the Grothendieck group that, for $g$ odd,
switches signs (naively exchanging branes and antibranes), and for
$g$ even, leaves the Grothendieck group essentially invariant.

Thus, if we apply Fourier-Mukai transform twice,
then we do not get precisely the same element of the Grothendieck
group we started with, but rather an element differing by a sign.
Thus, we can clearly
identify Fourier-Mukai transforms with T-duality only up to
a ${\bf Z}_2$.  As mentioned earlier, this is not a fundamental
difficulty, but merely reflects the fact that we have defined the
Fourier-Mukai transform symmetrically with respect to a torus and
its dual, rather than with sign asymmetries that are often introduced.

\subsection{Non-W.I.T. sheaves}

Earlier we gave the definition of W.I.T. sheaves, and noted
that for W.I.T. sheaves, Fourier-Mukai transforms simplify
greatly -- their action becomes well defined on individual
W.I.T. sheaves, one does not need the full technology of
derived categories and/or Grothendieck groups.
In prior physics literature on Fourier-Mukai transforms,
all sheaves were typically assumed to be W.I.T., for precisely
this reason.
Unfortunately, not all the coherent sheaves that one
would like to study are W.I.T. -- not even all supersymmetric D-brane
vacua are W.I.T. -- and for the more general
case one needs the more sophisticated methods reviewed in this
paper.  In this section we shall work through an example of a non-W.I.T.
sheaf, and speak to the relationship between the
W.I.T. condition and Mumford-Takemoto semistability.

First, let us construct an easy explicit example of a non-W.I.T. sheaf.
Consider a sheaf ${\cal E} \oplus {\cal T}$ on $T^2$,
where ${\cal E}$ is a W.I.T. rank $N$ bundle of $c_1 = 0$, and ${\cal T}$
is a torsion sheaf supported at $N'$ points on $T^2$.
It is easy to check that this sheaf is not W.I.T.
As described earlier, the Fourier-Mukai transform of ${\cal E}$
is $\ominus \hat{ {\cal E} } = \ominus R^1 \pi_{2 *} \left( 
{\cal P} \otimes \pi_1^* {\cal E} \right) \in K'_0( \hat{T}^2 )$,
and the Fourier-Mukai transform of ${\cal T}$ is a rank $N'$
bundle $\hat{ {\cal T} }$ on $\hat{T}^2$.
Thus, the Fourier-Mukai transform of the sheaf ${\cal E} \oplus {\cal T}$  
is the virtual sheaf $\hat{ {\cal T} } \ominus \hat{ {\cal E} }
\in K'_0(\hat{T}^2)$.
In other words, the Fourier-Mukai transform of a non-W.I.T. sheaf
is not an honest sheaf, but rather some general element of
the Grothendieck group of coherent sheaves (or, depending on the
reader's preference, the derived category of coherent sheaves)
on the dual algebraic torus.

What is the physics buried in the mathematical example above?
The coherent sheaf ${\cal E} \oplus {\cal T}$ cannot be a supersymmetric
vacuum configuration -- it corresponds to non-dissolved D0-branes
inside D2-branes.  The Fourier-Mukai transformation takes
this non-supersymmetric configuration, involving only branes,
to another non-supersymmetric configuration, but (at least naively)
involving both branes and antibranes.  At first blush it seems very
surprising that T-duality could map a configuration of only branes to one
involving both branes and antibranes.  However, on both sides of the
duality we have a nonsupersymmetric configuration, and perhaps more
importantly, it is not clear how to distinguish a configuration of
D0-branes and D2-antibranes from a configuration of D0- and D2-branes.
We shall return to this issue after making a closer examination of the
W.I.T. condition.

The reader might well ask, what is the precise relationship
between Mumford-Takemoto semistability and the W.I.T. condition?
For example, the reader may be tempted to suspect that 
supersymmetric brane vacua are W.I.T. and so have easy 
Fourier-Mukai transformations, in other words, that
a locally-free
sheaf that is Mumford-Takemoto semistable (and therefore satisfies
necessary conditions for a supersymmetric vacuum for a brane)
must be W.I.T.
Unfortunately this does
not seem to be the case in general \cite{huyb,mac,thom}.

Under what circumstances is a torsion-free, Mumford-Takemoto semistable
sheaf ${\cal E}$ also W.I.T.?  First, let us specialize to 
the case that ${\cal E}$ is Mumford-Takemoto stable, not just semistable,
and that\footnote{It is interesting that W.I.T. and stability
of a torsion-free sheaf ${\cal E}$ correlate
somewhat more naturally when $\mbox{det } {\cal E}$ is trivial;
one is tempted to wonder if there is any connection to the fact
that overall $U(1)$'s decouple from $U(N)$ in the AdS/CFT
correspondence (see, for example, \cite{edofer}).} $c_1({\cal E}) = 0$.
In this case, ${\cal E}$ can have no (holomorphic) sections,
as such a section would make ${\cal O}$ a subsheaf of the same
slope as ${\cal E}$, whose existence would contradict Mumford-Takemoto
stability.  Similarly, if ${\cal L}$ is any flat line bundle,
then ${\cal E} \otimes {\cal L}$ cannot have a section,
as the section would define ${\cal L}^{\vee}$ as a subsheaf,
and we would have the same contradiction as for ${\cal O}$.
Thus, for any flat line bundle ${\cal L}$, $H^0({\cal E} \otimes {\cal L})
= 0$, and so\footnote{In this subsection we shall be slightly
sloppy about computing right derived functors.  For a more detailed
examination of their properties, see for example 
\cite[section III.12]{hartshorne}.} 
$R^0 \pi_{2 *} \left( {\cal P} \otimes \pi_1^* {\cal E} \right) = 0$.
Now, for any torsion-free sheaf ${\cal E}$,
${\cal E}$ is Mumford-Takemoto stable if and only if ${\cal E}^{\vee}$
is also Mumford-Takemoto stable \cite[lemma 4.5]{friedbook},
consequently by Serre duality we have that on an $n$-(complex-)dimensional 
torus, $H^n({\cal E} \otimes {\cal L}) = 0$ for any flat line bundle
${\cal L}$ by the same arguments as above,
and so $R^n \pi_{2 *} \left( {\cal P} \otimes \pi_1^* {\cal E} \right) = 0$.

Thus, a torsion-free, Mumford-Takemoto stable sheaf on $T^4$ of $c_1 = 0$
is necessarily W.I.T.$_1$.  Unfortunately one does not get such statements
in greater generality.  For example, on higher-dimensional tori,
there is no good reason why a torsion-free, Mumford-Takemoto stable
sheaf of $c_1 = 0$ should be W.I.T., and in general we expect that they
will not be W.I.T.

So far in our discussion of the relationship between the W.I.T.
condition and Mumford-Takemoto stability, we have only spoken about
stable sheaves.
How would one deal with Mumford-Takemoto semistable sheaves
that are not stable?  After all, these can also satisfy the conditions for
a supersymmetric D-brane vacuum.  As noted in \cite{kcsub},
when using a properly semistable sheaf, physics sees a split sheaf
with stable factors.  Thus, questions regarding Fourier-Mukai transforms
and W.I.T. conditions for semistable sheaves can be reduced to questions
regarding direct sums of stable sheaves.

In general, therefore, there does not seem to be a simple relationship
between Mumford-Takemoto stability and the W.I.T. condition.
If we follow the usual wisdom that a ${\bf Z}_2$ subgroup of T-duality
is identified with Fourier-Mukai transformation, then 
one consequence is that T-duals of some supersymmetric D-brane vacua
naively involve both branes and antibranes.

Some care is required in interpreting Grothendieck group
elements, however.
A standard example from topological $K$-theory should make
possible subtleties more clear.  Let ${\cal E}$ be a $C^{\infty}$ vector
bundle on a $k$-dimensional manifold $M$, then there exists a rank $k$
$C^{\infty}$ bundle ${\cal F}$ such that ${\cal E} = 1 \ominus {\cal F}$,
where $1$ denotes the trivial rank $2k$ bundle
\cite[exercise 3.3e, p. 39]{husemoller}.
Ordinarily, following \cite{edktheory}, one would assume that
$1 \ominus {\cal F}$ was necessarily a non-supersymmetric configuration
of both branes and antibranes, but here we see that even without
tachyon condensation, sometimes a brane/antibrane configuration is
equivalent to a configuration of only branes\footnote{For a simpler
example of a brane/antibrane configuration equivalent to a configuration
of only branes without tachyon condensation, consider the 
$K$-theory element $\left( {\cal E} \oplus {\cal F} \right) \ominus
{\cal F}$.  This is clearly the same as the branes-only configuration
${\cal E}$, without tachyon condensation.  The example discussed
above is merely a more sophisticated version of this case.}.

Thus, even without tachyon condensation, sometimes naively nontrivial
elements of topological $K$-theory, and also Grothendieck groups,
are equivalent to trivial elements.  We suspect (though we have not
proven) that this is what is happening in Fourier-Mukai transformations
of non-W.I.T. supersymmetric D-brane vacua -- one gets a naively nontrivial
element of the Grothendieck group of coherent sheaves, which is subtly
equivalent to a configuration involving only branes.

\section{Conclusions}

In this paper we have argued that it can be
useful to consider D-brane charges in terms of Grothendieck groups
and, to a lesser extent, derived categories.  We began by
defining Grothendieck groups and listing some basic properties,
then briefly outlined derived categories and displayed a 
physically natural map from objects of a derived category to
a Grothendieck group of coherent sheaves.  We concluded with a
discussion of T-duality in terms of Fourier-Mukai transforms,
and argued that to understand the action of Fourier-Mukai
transforms even on general supersymmetric vacua, one needed
the technology of Grothendieck groups and derived categories.

Derived categories have previously entered the physics literature
through Kontsevich's mirror conjecture \cite{zaslow,kont},
in which mirror symmetry was conjectured to be realizable as
an equivalence of certain derived categories.
It would be interesting to see if any insight could be gained
by working instead with Grothendieck groups.
Perhaps Grothendieck groups would be more relevant for
the open-string mirror symmetry proposed in \cite{vafamir}. 

Derived categories might conceivably play a role in giving
a solid justification to certain proposed analogues of T-duality.
For example, in \cite{bjpsv}, a T-duality symmetry was conjectured 
that exchanged branes wrapped on general algebraic surfaces.
This hypothesized T-duality-analogue might conceivably be justified in terms
of an equivalence of derived categories on algebraic surfaces,
or even an automorphism of derived categories on a single\footnote{In
particular, it has been argued \cite{bondalorlovI,bondalorlovII} 
that a variety $X$ can be
more or less reconstructed from $D^b_c(X)$ if either its canonical
sheaf or its anticanonical sheaf is ample, so for some algebraic surfaces,
such as del Pezzo surfaces, the only possible T-duality-analogues of the form
proposed above would necessarily map the surface into itself.} algebraic
surface,
in which case one would presumably get not a ${\bf Z}_2$ subgroup of
some continuous family of symmetries, but only a (discrete)
${\bf Z}_2$ T-duality-analogue.  (A more specialized form
of this conjecture, given essentially for algebraic K3s, was stated in 
\cite{horioz}.)  One might even speculate that the existence of 
equivalences of derived categories of coherent sheaves on
distinct Calabi-Yau's \cite{macpap} might signal the existence
of some mirror-symmetry-analogue for branes, analogous to that
proposed in \cite{vafamir}.

The description of D-branes in terms of $K$ theory given
in \cite{edktheory} may also yield interesting new insights
via string-string duality.  For example, consider the duality
relating IIA compactified on $K3$ to a heterotic string on $T^4$.
A D2-brane wrapped on a curve in $K3$, for example, is
interpreted as a particle on the heterotic side.
What is the heterotic interpretation of a wrapped
D2 brane/antibrane pair?  Presumably the heterotic dual to such
a configuration is a massive heterotic string state
\cite{berggaber}.
Thus, the interpretation of D-branes in terms of $K$ theory
may give rise to a new geometric interpretation of massive
heterotic states, for example.

It is somewhat tempting to speculate that massive heterotic
string states may have, at least sometimes, an interpretation
in terms of topological $K$ theory or Grothendieck groups of
the space that the heterotic string is compactified on.  
In such an event, it would seem
likely that isomorphisms of derived categories on distinct
Calabi-Yau's \cite{macpap} may correspond immediately to some limit
of (0,2) mirror symmetry, in which all $B$-fields are turned off
and $\alpha'$ is small.

\section{Acknowledgements}

We would like to thank P.~Aspinwall, T.~Gomez, K.~Hori, P.~Horja,
A.~Knutson, D.~Morrison, 
B.~Pardon,
R.~Plesser, M.~Stern, and E.~Witten for useful conversations.

\appendix

\section{Notes on topological $K$-theory}

For more information on topological $K$-theory,
see for example \cite{ah,husemoller,karoubi,atiyah}.

Given a compact complex manifold $X$,
the group $K(X)$ is the free abelian group on
isomorphism classes of complex vector bundles on $X$,
modulo elements of the form $[{\cal E}_0 \oplus {\cal E}_1]
 - [{\cal E}_0] - [{\cal E}_1]$, where ${\cal E}_0$,
${\cal E}_1$ are complex vector bundles on $X$ and
$[{\cal E}]$ denotes the isomorphism class of ${\cal E}$.
Put another way, elements of $K(X)$ are ``virtual bundles''
of the form ${\cal E} \ominus {\cal F}$.
More generally, it is straightforward to see that
$K$ defines a contravariant functor from the
category of compact spaces to the category of abelian groups.

What is $K(\mbox{point})$ ?  A vector bundle on a point
is completely determined by its rank, so it should be
clear that $K(\mbox{point}) \cong {\bf Z}$.

The reduced $K$-ring of $X$, denoted $\tilde{K}(X)$,
is defined to be the kernel of the natural projection
$K(X) \rightarrow K(\mbox{point})$.

Let $X$ be a topological space, $Y$ some subset of $X$.
There exists a notion of relative K-theory,
denoted $K(X,Y)$, which is defined as
$K(X,Y) = \tilde{K}(X/Y)$.  (If $Y$ is a closed subset of $X$,
then $X/Y$ is essentially a space obtained by collapsing $Y$
down to a single point.  For more information see for example
\cite[section 0.2]{hatcher}.)  As the reader may well guess,
$K(X,\emptyset) = K(X)$.

Define the suspension of a topological space $X$, denoted
$SX$, to be the quotient of $X \times I$ (where $I = [0,1]$,
the unit interval) obtained by collapsing $X \times \{ 0 \}$
to one point and $X \times \{ 1 \}$ to another point.
For example, $S S^n = S^{n+1}$.

For $n$ positive, define 
$K^{-n}(X/Y) = \tilde{K}( S^n( X/Y ) )$.  In particular,
$K^{-1}(\mbox{point}) = 0$.

Bott periodicity is simply the statement that for any
compact Hausdorff space $X$, $K^n(X) \cong K^{n+2}(X)$.
Similarly, $K^n(X/Y) \cong K^{n+2}(X/Y)$.

We defined $K^n$ above for $n$ negative only;
however, by Bott periodicity we can now define
$K^n$ for arbitary integer $n$:
\begin{eqnarray*}
K^n(X) & = & K^0(X) \mbox{ for n even} \\
K^n(X) & = & K^{-1}(X) \mbox{ for n odd}
\end{eqnarray*}
and similarly for $K^n(X,Y)$, and so forth.

$K^1(X)$ has an alternative definition, described in
\cite[section II.3]{karoubi}.  Consider the category whose
objects are pairs
$({\cal E}, \alpha)$ where ${\cal E}$ is a bundle and
$\alpha: {\cal E} \rightarrow {\cal E}$ is an isomorphism,
and whose morphisms $({\cal E}, \alpha) \rightarrow ({\cal E}',
\alpha')$ are given by maps $h: {\cal E} \rightarrow {\cal E}'$
such that the following commutes:
\begin{displaymath}
\begin{array}{ccc}
{\cal E} & \stackrel{\alpha}{\longrightarrow} & {\cal E} \\
h \downarrow & & \downarrow h \\
{\cal E}' & \stackrel{\alpha'}{\longrightarrow} & {\cal E}'
\end{array}
\end{displaymath}
Define the sum of two objects $({\cal E}, \alpha)$ and
$({\cal E}', \alpha')$ to be $( {\cal E} \oplus {\cal E}',
\alpha \oplus \alpha')$.
Define a pair $({\cal E}, \alpha)$ to be elementary if
$\alpha$ is homotopic to the identity within automorphisms of
${\cal E}$.  Now we finally have the definitions in hand
to define $K^1(X)$.  Define $K^1(X)$ to be the free abelian group
on objects in the category, modulo the equivalence relation
$({\cal E}, \alpha) \sim ({\cal E}', \alpha')$ if and only
if there exist elementary pairs $({\cal F}, \beta)$ and $({\cal F}', \beta')$
such that $({\cal E}, \alpha) + ({\cal F}, \beta) \cong
({\cal E}', \alpha') + ({\cal F}', \beta')$.

For notational purposes, let $[{\cal E}, \alpha]$ denote
the equivalence class of the pair $({\cal E}, \alpha)$ 
in $K^1(X)$.  Then it can be shown that
$[{\cal E}, \alpha \circ \beta] = [{\cal E}, \alpha] +
[{\cal E}, \beta]$.

It is possible to define a product on elements of
$K$ theory; it has the properties
\begin{eqnarray*}
K^0(X) \cdot K^0(X) & \subseteq & K^0(X) \\
K^0(X) \cdot K^1(X) & \subseteq & K^1(X) \\
K^1(X) \cdot K^1(X) & \subseteq & K^0(X) 
\end{eqnarray*}

By this point the reader has no doubt noticed the similarity
between the groups $K^n(X)$ and $H^n(X)$.  In fact,
$K$ theory is an example of a ``generalized'' cohomology
theory.  More precisely, cohomology theories can be 
defined axiomatically \cite{es}, and $K$ theory satisfies
all the axioms for a cohomology theory except one (the
dimension axiom).

\end{document}